\documentclass[showkeys,aps,showpacs,amssymb,prd]{revtex4}

\usepackage{appendix}
\usepackage{amsbsy}
\usepackage{amsfonts}
\usepackage{amsmath}
\usepackage{amssymb}
\usepackage{amsthm}

\usepackage{graphicx}
\usepackage{ifthen}
\usepackage{textcomp}

\newcounter{multi} \newcounter{multa}

\newcounter{faki} \newcounter{faka}

    \newtheorem{theorem}{Theorem}

\theoremstyle{definition} 

\theoremstyle{remark} 

\newcommand{\C}{\mathbb{C}}
\newcommand{\CP}{\mathbb{CP}}
\newcommand{\SP}{\mathbb{S}}
\newcommand{\WP}{\mathbb{WP}}
\newcommand{\wt}{\widetilde}
\newcommand{\eqn}{eqn.~\eqref}

\newcommand{\beqa}{\begin{eqnarray}}
\newcommand{\beq}{\begin{equation}}
\newcommand{\eeqa}{\end{eqnarray}}
\newcommand{\eeq}{\end{equation}}

\newcommand{\bo}{\boldsymbol}

\newcommand*{\longhookrightarrow}{\ensuremath{\lhook\joinrel\relbar\joinrel\rightarrow}}

\newcommand{\lra}{\longrightarrow}
\newcommand{\fkM}{\mathfrak{M}}
\newcommand{\fkMt}{\mathfrak{M}_{\text{tot}}}

\def\det {\mathop{\rm det}\nolimits}

\def\Jac {\mathop{\rm Jac}\nolimits}

\def\RR{\mathbb{R}}
\def\bbf{{\bf f}}
\def\bc{{\bf c}}
\def\bx{{\bf x}}
\def\bby{{\bf y}}
\def\blm{\boldsymbol \eta}
\def\bs{{\bf s}}

\def\detfA{{\rm det}\left({\rm Jac}\,\bbf^{A_n}_{\bc}\right)}

\def\fA{{\bbf}^{A_n}_{\bc}}
\def\fD{{\bbf}^{D_{n}^{\pm}}_{\bc}}

\def\fEsix{{\bbf}^{E_6}_{\bc}}
\def\fEseven{{\bbf}^{E_7}_{\bc}}
\def\fEeight{{\bbf}^{E_8}_{\bc}}

\def\blm{\boldsymbol \eta}

\begin{document}

\title{Orbifolds, the $A, D, E$ Family of Caustic Singularities, and Gravitational Lensing}

\author{A. B. Aazami}\email{aazami@math.duke.edu}\affiliation{Department of
Mathematics, Duke University, Science Drive, Durham, NC 27708}

\author{A. O. Petters}\email{petters@math.duke.edu}\affiliation{Departments of
Mathematics and Physics, Duke University, Science Drive, Durham, NC 27708}

\author{J. M. Rabin}\email{jrabin@math.ucsd.edu}\affiliation{Department of Mathematics, University of California, San Diego, CA 92093}

\begin{abstract}
We provide a geometric explanation for the existence of magnification relations for the $A_n~(n \geq 2), D_n~(n \geq 4), E_6, E_7, E_8$ family of caustic singularities, which were established in recent work.  In particular, it was shown that for families of general mappings between planes exhibiting any of these caustic singularities, and for any non-caustic target point, the total signed magnification of the corresponding pre-images vanishes.  As an application to gravitational lensing, it was also shown that, independent of the choice of a lens model, the total signed magnification vanishes for a light source anywhere in the four-image region close to elliptic and hyperbolic umbilic caustics.  This is a more global and higher-order analog of the well-known fold and cusp magnification relations.  We now extend each of these mappings to weighted projective space, which is a compact orbifold, and show that magnification relations translate into a statement about the behavior of these extended mappings at infinity.  This generalizes multi-dimensional residue techniques developed in previous work, and introduces weighted projective space as a new tool in the theory of caustic singularities and gravitational lensing.
\end{abstract}

\maketitle

\section{Introduction}
\label{Introduction}
In a series of recent papers (Aazami \& Petters 2009, 2010 \cite{Aazami-Petters,Aazami-Petters2,Aazami-Petters3}), the authors established a universal magnification theorem for all caustic singularities classified by the $A_n~(n \geq 2), D_n~(n \geq 4), E_6, E_7, E_8$ family of Coxeter-Dynkin diagrams.  In particular, it was shown in \cite{Aazami-Petters2,Aazami-Petters3} that to each caustic singularity in this family is associated a magnification relation of the form
\beq
\label{magrel}
\sum_{i=1}^{n}\fkM_i = 0.
\eeq
Such relations are not only of mathematical interest, but are important in gravitational lensing as well.  In particular, slices of the big caustics of some of these singularities give rise to caustic metamorphoses that occur in gravitational lensing (e.g., Blandford 1990 \cite{Blandford90}, Schneider, Ehlers, \& Falco 1992 \cite{Sch-EF}, Petters 1993 \cite{Petters93}, and Petters et al. 2001 \cite[Chaps. 7, 9]{Petters}).  Moreover, in \cite{Aazami-Petters} it was shown that, independent of the choice of a lens model, such magnification relations hold for a source lying anywhere in the four-image region near elliptic umbilic $(D_4^{-})$ and hyperbolic umbilic $(D_4^{+})$ caustics.  It was further shown how the hyperbolic umbilic magnification relation could be used for substructure studies of four-image lens galaxies, analogous to the well-known fold $(A_2)$ and cusp $(A_3)$ magnification relations (e.g., Blandford \& Narayan 1986 \cite{Blan-Nar}, Schneider \& Weiss 1992 \cite{Sch-Weiss92}, Zakharov 1995 \cite{Zakharov}, Mao \& Schneider 1998  \cite{Mao-Sch}, \cite{Petters}, Keeton, Gaudi \& Petters 2003 and 2005 \cite{KGP-cusps,KGP-folds}).  Using Lefschetz fixed point theory, Werner 2007 \& 2009 \cite{Werner,Werner2} has recently shown that such relations are in fact topological invariants.

The proof of \eqn{magrel} in \cite{Aazami-Petters2,Aazami-Petters3} was algebraic, making repeated use of the Euler trace formula.  The aim of the present paper is to give a {\it geometric} explanation for the existence of such relations.  We do so by generalizing the multi-dimensional residue technique developed by Dalal \& Rabin 2001 \cite{Dalal-Rabin}.  Their procedure was as follows.  Each caustic singularity appearing in Arnold's classification gives rise (through its gradient) to a mapping between planes.  Complexifying, and using homogeneous coordinates, one can extend these mappings to the complex projective plane $\CP^2$.  Next, the magnifications $\fkM_i$ appearing in \eqn{magrel} are realized as residues of a certain meromorphic two-form.  By the Global Residue Theorem (Griffiths \& Harris 1978 \cite{Griffiths-Harris}), the sum of these residues, which reside in affine space, is precisely equal to minus the sum of the residues at infinity.  A magnification relation such as \eqn{magrel} is thus transformed into a statement about the behavior of these (extended) mappings at infinity in $\CP^2$.

Ideally, if the right-hand side of a magnification relation is identically zero, one would like for there to be {\it no} residues at infinity.  For the $A_n~(n \geq 2), D_n~(n \geq 4), E_6, E_7, E_8$ family of caustic singularities, however, this is not always the case.  The way around this is to consider extensions into spaces other than $\CP^2$, namely, the so-called weighted projective spaces $\WP(a_0,a_1,a_2)$.  These are compact orbifolds which have recently come into prominence in string theory (see, e.g., Adem et al. 2007 \cite{Orbistrings}).  We show that one can extend each mapping associated to a caustic singularity to a particular weighted projective space such that there will be {\it no} residues at infinity.  Magnification relations such as \eqn{magrel} are then immediately explained.

The outline of this paper is as follows.  In Section~\ref{ADE} we give a brief overview of the $A, D, E$ family of caustic singularities.  In Section~\ref{residuethm} we outline the residue method in \cite{Dalal-Rabin} and show how the Global Residue Theorem can be applied to compact orbifolds.  For convenience, we restate in Section~\ref{Theorem} the main theorem in \cite{Aazami-Petters,Aazami-Petters2,Aazami-Petters3}.  For readers unfamiliar with the basic properties of orbifolds and with weighted projective space in particular, we provide a review in Appendix~\ref{WP}.  The residue proof itself is given in Appendix~\ref{proof}, including the lensing map cases established in \cite{Aazami-Petters}.

\section{Caustic Singularities of the $A, D, E$ family}
\label{ADE}

In what follows, we use the notation and terminology of \cite{Aazami-Petters,Aazami-Petters2,Aazami-Petters3}.  Consider a smooth $k$-parameter family $F_{{\bc},{\bs}}(\bx)$ of functions on an open subset of $\mathbb{R}^2$ that induces a smooth $(k-2)$-parameter family of mappings $\bbf_{\bc}(\bx)$ between planes ($k \geq 2)$.  One uses $F_{\bc,\bs}$ to construct a {\it Lagrangian submanifold} that is projected into the space $\{{\bc},{\bs}\} = \mathbb{R}^{k-2} \times \mathbb{R}^{2}$.  The caustics of $\bbf_{\bc}$ will then be the critical values of the projection  (e.g.,  Golubitsky \& Guillemin 1973 \cite{Gol-G}, Majthay 1985 \cite{Majthay}, Castrigiano \& Hayes 1993 \cite{C-Hayes}, and \cite[pp. 276-86]{Petters}).  These projections are called {\it Lagrangian maps}, and they are differentiably equivalent to  $\bbf_\bc$.  Arnold classified all stable simple Lagrangian map-germs of $n$-dimensional Lagrangian submanifolds by their generating family $F_{\bc,\bs}$ (\cite{Arnold73}, Arnold, Gusein-Zade, \& Varchenko I 1985 \cite[p. 330-31]{AGV1}, and \cite[p. 282]{Petters}).  In the process he found a connection between his classification and the Coxeter-Dynkin diagrams of the simple Lie algebras of types $A_n~(n \geq 2), D_n~(n \geq 4), E_6, E_7, E_8$.  This classification is shown in Table~\ref{table1}.  (The classification of the elementary catastrophes, for codimension less than 5, was determined by Ren\'e Thom in the 1960s.)

\begin{table}
\footnotesize
\centering
\vskip 6pt
\begin{tabular}{| c | c |}
\hline
& \\
& $F_{\bc,\bs}(x,y) =  \pm x^{n+1} \pm y^2 + c_{n-1}x^{n-1} + \cdots + c_3x^3 + s_2x^2 - s_1x \pm s_2y$ \\
~~~~~~{\it \large{A}}$_{n}~~~(n \geq 2)$~~~~~~ & \\
& ${\bf f}_{\bc}(x,y) = \left(\pm(n+1)x^{n} + (n-1)c_{n-1}x^{n-2} + \cdots + 3c_3x^2 - 4yx~,~\mp2y\right)$ \\
& \\
\hline
& \\
& $F_{\bc,\bs}(x,y) = x^2y \pm y^{n-1} + c_{n-2}y^{n-2} + \cdots + c_2y^2 - s_2y - s_1x$ \\
~~~~~~{\it \large{D}}$_n~~~(n \geq 4)$~~~~~~ & \\
& ${\bf f}_{\bc}(x,y) = \left(2xy~,~x^2\pm(n-1)y^{n-2} + (n-2)c_{n-2}y^{n-3} + \cdots + 2c_2y\right)$ \\
& \\
\hline
& \\
& $F_{\bc,\bs}(x,y) = x^3 \pm y^4 + c_3xy^2 + c_2y^2 + c_1xy - s_2y - s_1x$ \\
~~~~~~{\it \large{E}}$_6$~~~~~~ & \\
& ${\bf f}_{{\bo c}}(x,y) = \left(3x^2 + c_3y^2 +c_1y~,~\pm4y^3 + 2c_3xy + 2c_2y + c_1x\right)$ \\
& \\
\hline
& \\
& $F_{\bc,\bs}(x,y) = x^3 + xy^3 + c_4y^4 + c_3y^3 + c_2y^2 + c_1xy - s_2y - s_1x$ \\
~~~~~~{\it \large{E}}$_7$~~~~~~ & \\
& ${\bf f}_{{\bo c}}(x,y) = \left(3x^2 + y^3 + c_1y~,~3xy^2 + 4c_4y^3 + 3c_3y^2 + 2c_2y + c_1x\right)$ \\
& \\
\hline
& \\
& $F_{\bc,\bs}(x,y) = x^3 + y^5 + c_5xy^3 + c_4xy^2 + c_3y^3 + c_2y^2 + c_1xy - s_2y - s_1x$ \\
~~~~~~{\it \large{E}}$_8$~~~~~~ & \\
& ~~~~${\bf f}_{{\bo c}}(x,y) = \left(3x^2 + c_5y^3 + c_4y^2 + c_1y~,~5y^4 + 3c_5xy^2 + 2c_4xy + 3c_3y^2 + 2c_2y + c_1x\right)$~~~~ \\
& \\
\hline
\end{tabular}
\caption{For each type of Coxeter-Dynkin diagram listed, indexed by $n$, the second column shows the corresponding universal local forms of the smooth $(n-1)$-parameter family of general functions $F_{{\bo c},\bs}$, along with their $(n-3)$-parameter family of induced general maps $\bbf_{\bo c}$ between planes.  This classification is due to Arnold 1973 \cite{Arnold73}.}
\label{table1}
\end{table}


The $\bbf_{\bc}$ shown in Table~\ref{table1} are obtained from their corresponding $F_{\bc,\bs}$ by taking its gradient with respect to $\bx$ and setting it equal to zero: grad$(F_{\bc,\bs})(\bx) = 0$.  This equation is then rewritten in the form $\bbf_{\bc}(\bx) = \bs$.  We call $\bx \in \RR^2$ a {\it pre-image} of the {\it target point} $\bs \in \RR^2$ if $\bbf_{\bc}(\bx) = \bs$.  Equivalently, this will be the case if and only if $\bx$ is a critical point of $F_{\bc,\bs}$ (relative to a gradient in $\bx$).  Next, we define the {\it magnification} $\fkM(\bx;\bs)$ at a critical point $\bx$ of the family $F_{{\bc}, \bs}$ by the reciprocal of the Gaussian curvature at the point $(\bx,F_{{\bc},\bs}(\bx_i))$ in the graph of $F_{{\bc},\bs}$:
\beq
\label{Gauss}
\fkM({\bx; \bs}) \equiv \frac{1}{{\rm Gauss}(\bx,F_{{\bc},\bs}(\bx))}\cdot\nonumber
\eeq
This makes it clear that the magnification invariants are {\it geometric} invariants.  In addition, since ${\rm Gauss}(\bx,F_{{\bc},\bs}(\bx)) = \text{det(Hess}\,F_{\bc,\bs})(\bx)$, and since each $\bbf_{\bc}$ in Table~\ref{table1} satisfies $\text{det(Jac}\,\bbf_{\bc}) = \text{det(Hess}\,F_{\bc,\bs})$, we can also express the magnification in terms of $\bbf_{\bc}$:
\beq
\label{Jac}
\fkM (\bx; \bs) = \frac{1}{\det(\Jac \bbf_{\bc})(\bx)}\cdot
\eeq
If $\bs$ has a total of $n$ pre-images $\bx_i$, then the {\it total signed magnification of $\bs$} is defined to be
\beq
\label{tot}
\fkMt(\bs) \equiv \sum_{i=1}^{n} \fkM(\bx_i;\bs).
\eeq
{\it Critical points} of $\bbf_{\bc}$ are those $\bx$ for which ${\rm det}({\rm Jac}\,\bbf_{\bc})(\bx) = 0$.  Generically, the locus of critical points forms curves called {\it critical curves}.  The target $\bbf_{\bc}(\bx)$ of a critical point $\bx$ is called a {\it caustic point}.  These typically form curves, but could be isolated points.  Varying ${\bc}$ causes the caustic curves to evolve.  This traces out a caustic surface, called a {\it big caustic,} in the $k$-dimensional space $\{\bc,\bs\} = \RR^{k-2} \times \RR^{2}$.  In particular, our magnification relations involve only non-caustic target points.

In the context of gravitational lensing, the family of maps $F_{\bc,\bs}$ is replaced by a family of time delay functions $T_{\bc,\bs}$.  To that end, consider a smooth $k$-parameter family of time delay functions $T_{\bc,\bs}(\bx)$ which induces an ($k-2$)-parameter family of lensing maps $\blm_{\bc}$.  The {\it universal, quantitative} form of the lensing map can be derived in a neighborhood of a caustic using rigid coordinate transformations and Taylor expansions, using appropriate constraint equations (see \cite[Chap.~6]{Sch-EF} and \cite{Aazami-Petters}).  For the elliptic umbilic $(D_4^{-})$ and the hyperbolic umbilic $(D_4^{+})$, these are shown in Table~\ref{table2}.  The definitions of magnification, critical points, and caustic points are the same as in the generic case.  However, we will follow the notational convention in \cite{Aazami-Petters} and use the symbol $\mu$ to denote the magnification in the lensing case.

\section{Multi-Dimensional Residue Techniques on Compact Orbifolds}
\label{residuethm}

The essence of the residue method developed in \cite{Dalal-Rabin} is to express the magnification $\fkM(\bx;\bs)$ in \eqn{Jac} as the residue of a meromorphic two-form defined on compact projective space $\CP^2$.  We summarize the procedure here, in the context of the general mappings $\bbf_\bc$ shown in Table~\ref{table1}; consult \cite{Dalal-Rabin} for a detailed treatment, including applications to realistic lens models in gravitational lensing.

\begin{table}[tbh]
\centering
\vskip 6pt
\begin{tabular}{| c | c |}
\hline
& \\
& \hspace{.65 in}$T_{c,\bs}(x_1,x_2) = {1 \over 2}{\bs}^{2}-{\boldsymbol {\rm x \cdot
s}}+{1 \over 3}x_1^{3}-x_1x_2^{2}+2cx_2^{2}\hspace{.65 in}$\\
Elliptic umbilic $(D_4^{-})$ & \\ 
& $\blm_{c}(x_1,x_2)=\left(x_1^{2}-x_2^{2}\ ,\
-2x_1x_2+4cx_2\right)$ \\
& \\
\hline
& \\
& $T_{c,\bs} (x_1,x_2) = {1 \over 2}{\bs}^{2}-{\boldsymbol {\rm x \cdot s}}+{1 \over 3}(x_1^{3}+x_2^{3})+2cx_1x_2$ \\
$~~~~~$Hyperbolic umbilic $(D_4^{+})~~~$& \\
& $\blm_{c}(x_1,x_2)=\left(x_1^{2}+2cx_2\ ,\ x_2^{2}+2cx_1\right)$ \\
& \\
\hline
\end{tabular}
\caption{For the two caustic singularities listed, the second column shows the corresponding universal local forms of the smooth three-parameter family of time delay functions $T_{c,\bs}$, along with their one-parameter family of lensing maps $\blm_c$.  The parameter $c$ will represent some physical input, such as the source redshift.  For a derivation, see Schneider, Ehlers, \& Falco 1992 \cite[Chap.~6]{Sch-EF}.}
\label{table2}
\end{table}

Let $\bbf_\bc$ be any mapping shown in Table~\ref{table1}, with a given pre-image $\bx = (x,y)$ of a non-caustic target point $\bs = (s_1,s_2)$.  Let $f_\bc^{(1)}(x,y)$ and $f_\bc^{(2)}(x,y)$ denote the two components of $\bbf_\bc$, with degrees $d_1$ and $d_2$, respectively.  We can then express the pre-image $\bx$ as a common root of the following two polynomials:
\beq
\label{P1}
P_1(x,y) \equiv f_\bc^{(1)}(x,y) - s_1\hspace{.25 in},\hspace{.25 in}P_2(x,y) \equiv f_\bc^{(2)}(x,y) - s_2.
\eeq
Note that 
\beq
\label{J}
J(\bx) \equiv \text{det}\!\left[\begin{array}{cc} \partial_xP_1 & \partial_yP_1 \\ \partial_xP_2 & \partial_yP_2 \\ \end{array}\right] = \fkM(\bx; \bs)^{-1}.\nonumber
\eeq
In particular, $J(\bx) \neq 0$ because $\bs$ is a non-caustic target point.  Now treat the pre-image coordinates $\bx = (x,y)$ as complex variables, so that $\bx \in \mathbb{C}^2$, and consider the following meromorphic two-form defined on $\mathbb{C}^2$:
\beq
\label{form}
\omega = \frac{dx\,dy}{P_1(x,y)P_2(x,y)}\cdot\nonumber
\eeq
At points where $J \neq 0$, the residue of $\omega$ is given by
\beq
\label{res}
{\rm Res}\,\omega = \frac{1}{J(x,y)} = \fkM(\bx;\bs).
\eeq
Thus we have expressed the magnification $\fkM(\bx;\bs)$ as the residue of a meromorphic two-form defined on $\mathbb{C}^2$.  Next, since $\mathbb{C}^2$ can be viewed as the affine piece of $\CP^2$, changing to homogeneous coordinates $[X:Y:U]$ with $x = X/U$ and $y = Y/U$ extends the two polynomials $P_i(x,y)$ to $\CP^2$:
\beq
\label{P2}
P_1(X,Y,U)_{\text{hom}} \equiv U^{d_1}f_\bc^{(1)}(X/U,Y/U) - s_1U^{d_1}\hspace{.25 in},\hspace{.25 in}P_2(X,Y,U)_{\text{hom}} \equiv U^{d_2}f_\bc^{(2)}(X/U,Y/U) - s_2U^{d_2}.
\eeq
Affine space corresponds to $U = 1$, in which case we recover \eqn{P1}.  We can similarly extend 
$\omega$ to a form on $\CP^2$, still denoted $\omega$, that is homogeneous of degree zero:
\beq
\label{omega}
\omega = {d(X/U) d(Y/U) \over P_1(X/U,Y/U) P_2(X/U,Y/U)} = 
{U^{d_1+d_2-3} (U dX dY - X dU dY - Y dX dU) \over P_1(X,Y,U)_{\text{hom}} P_2(X,Y,U)_{\text{hom}}}.
\eeq

Since $\CP^2$ is a compact smooth manifold, the Global Residue Theorem states that the sum of the residues of any meromorphic form, such as $\omega$, on $\CP^2$, is identically zero.  Since all the poles of $\omega$ in affine space correspond to pre-images of $\bbf_\bc$ and vice-versa, the sum of their residues is the total signed magnification $\fkMt(\bs)$ given by \eqn{tot}.  The Global Residue Theorem thus states that $\fkMt(\bs)$ is precisely equal to minus the sum of the residues of $\omega$ at infinity $(U = 0)$.  This is the fundamental explanation of magnification relations established in \cite{Dalal-Rabin}: {\it the total signed magnification corresponding to a non-caustic target point of a mapping $\bbf_\bc$ reflects the behavior of $\bbf_\bc$ at infinity when it is extended to $\CP^2$.}  So in particular, if the homogeneous polynomials in \eqn{P2} have {\it no} common roots at infinity, then $\omega$ has no poles at infinity and thus no residues at infinity, and we can immediately conclude that $\fkMt(\bs) = 0$.  If there {\it are} common roots at infinity, then $\omega$ will have poles at infinity and their residues will have to be computed.  In \cite{Dalal-Rabin} a procedure for doing this was outlined and used to uncover magnification relations corresponding to a variety of lens models in gravitational lensing.  Such residues in general cannot be computed via \eqn{res}, because zeros at infinity may not satisfy $J \neq 0$.  Instead they are computed using the Leray residue formula, details of which can be found in \cite{Dalal-Rabin}.  Note in any case that the mappings are always extended to the compact smooth manifold $\CP^2$.  It is precisely this extension that we generalize here.

Given the simple form of the magnification relations
$$
\fkMt(\bs) = \sum_{i=1}^{n} \fkM_i = 0,
$$
one would expect there to be {\it no} common roots at infinity and thus no residue to calculate.  This, however, is not the case for some of the induced general mappings $\bbf_\bc$ shown in Table~\ref{table1}.  Take for example the $D_5$ caustic singularity (the parabolic umbilic), whose mapping is
\beq
\label{D5}
{\bf f}_{\bc}(x,y) = \left(2xy~,~x^2\pm4y^{3} + 3c_{3}y^{2} + 2c_2y\right) = (s_1,s_2),
\eeq
where once again $(s_1,s_2)$ is a non-caustic target point.  Extending this mapping via homogeneous coordinates into $\CP^2$ leads to the following two polynomials, as in \eqn{P2}:
\beq
\label{D52}
\text{$\bbf_\bc$ homogenized in $\CP^2~\Longrightarrow$}~~
\left\{
\begin{array}{ll}
2XY - s_1U^2\\
X^2U\pm4Y^{3} + 3c_{3}Y^{2}U + 2c_2YU^2 -s_2U^3.
\end{array}
\right.
\eeq
In affine space $(U=1)$ we recover \eqn{D5}.  At infinity $(U = 0)$, however, there is one nonzero common root, namely the point $[1:0:0]$ (recall that in homogeneous coordinates $[X:Y:U] = [X':Y':U'] \iff$ there is a nonzero $\lambda \in \mathbb{C}$ with $X = \lambda X',~Y = \lambda Y',~U = \lambda U'$; recall also that $[0:0:0] \notin \CP^2$).  The residue of $\omega$ at this point will therefore have to be computed.  This is not difficult, and the residue will be zero (as expected).  Nevertheless, this leads to the following question: can we find an extension of \eqn{D5} to a compact space {\it other} than $\CP^2$ that ensures there will be {\it no} common roots at infinity?  The answer is yes: consider the weighted projective space $\WP(3,2,1)$ (see Appendix~\ref{WP}).  In homogeneous coordinates, the difference between $\WP(3,2,1)$ and $\CP^2$ is the following: because of the action given by \eqn{action2},
$$
z \cdot (w_0,w_1,w_2) = (z^{3}w_0,z^{2}\,w_1,z\,w_2),
$$
the variables $w_0, w_1$ in $\WP(3,2,1)$ now have {\it weights} associated with them.  As a result, the relationship between homogeneous and affine coordinates is now given by
\beq
x= \frac{X}{U^3}\hspace{.25 in},\hspace{.25 in}y = \frac{Y}{U^2}\cdot \nonumber
\eeq
Extending \eqn{D5} to $\WP(3,2,1)$ thus gives the following two polynomials, which are different from those in \eqn{D52}:
\beq
\label{D53}
\text{$\bbf_\bc$ homogenized in $\WP(3,2,1)~\Longrightarrow$}~~
\left\{
\begin{array}{ll}
2XY - s_1U^5\\
X^2\pm4Y^{3} + 3c_{3}Y^{2}U^2 + 2c_2YU^4 - s_2U^6.
\end{array}
\right.
\eeq
Once again in affine space $(U = 1)$ we recover \eqn{D5}.  The situation at infinity $(U = 0)$, however, is now decidedly better, because the only common root of \eqn{D53} at infinity is the point $[0:0:0]$, which of course is not a point in $\WP(3,2,1)$.  We have therefore found an extension in which there are no roots at infinity.  Moreover, the only singularities of the orbifold $\WP(3,2,1)$ occur at infinity, because $U$ has weight 1.  In other words, the only $z \in \SP^1$ that satisfies $z\,w_2 = w_2$ for $w_2 \neq 0$ is $z = 1$.  In fact the only singular points of $\WP(3,2,1)$ are $[1:0:0]$ and $[0:1:0]$, with local groups isomorphic to $\mathbb{Z}/3\mathbb{Z}$ and $\mathbb{Z}/2\mathbb{Z}$, respectively.  Thus there are no singular points in affine space, where the pre-images reside.

As we will see in Appendix~\ref{proof}, an extension such as that in \eqn{D53}, in which there are no common roots at infinity, can be obtained for {\it all} the caustic singularities of the $A, D, E$ family.  Each such weighted projective space will be of the form $\WP(a_0,a_1,1)$, so it will have no singular points in affine space. The common roots lie in the affine subset $U=1$ which is nonsingular and simply $\C^2$.  The vanishing of the total magnification in $\C^2$ then follows from the orbifold version of the Global Residue Theorem.

The Global Residue Theorem as presented in \cite{Griffiths-Harris} applies to compact smooth manifolds only. The extension to compact orbifolds is Remark 4.10 of Cattani, Cox, \& Dickenstein 1997 
\cite{Cattani-Cox}. However, Cattani, Dickenstein, \& Sturmfels 1996 \cite{Cattani-Dickenstein} give a useful statement adapted to our context of weighted projective spaces (Corollary 1.18). Consider a generalization of the form $\omega$ in $\C^2$,
\beq
\label{omega-h}
\omega = \frac{h(x,y) \,dx\,dy}{P_1(x,y)P_2(x,y)},
\eeq
where $h(x,y)$ is a polynomial. Such a form can occur in the computation of total magnification when $x,y$ are non-rectangular coordinates for the pre-images, or more generally in computing moments of the magnification. Then $\omega$ has no residue at infinity in $\WP(a_0,a_1,1)$ whenever it has negative degree, that is, when
\beq
\label{neg-deg}
\deg h < \deg P_1 + \deg P_2 - a_0 - a_1.
\eeq
In this statement it is understood that all degrees are weighted, so that $\deg x = a_0$ and 
$\deg y = a_1$.
When the degree of $\omega$ is nonnegative, a simple algorithm is given in \cite{Cattani-Dickenstein} to compute the residue at infinity.

\section{Restatement of Main Theorem}
\label{Theorem}
For convenience we restate the Theorem established in \cite{Aazami-Petters,Aazami-Petters2,Aazami-Petters3}.
\begin{theorem}
\label{theorem-main}
For any of the universal, smooth $(n-1)$-parameter families of general functions $F_{{\bc},\bs}$ {\rm(}or induced general mappings $\bbf_\bc${\rm)} in Table~\ref{table1}, and for any non-caustic target point $\bf s$ in the indicated region, the following results hold for the magnification $\fkM_i \equiv  \fkM({\bx_i; \bs})$:

\begin{enumerate}
\item $A_n~(n \geq 2)$ obeys the magnification relation in the $n$-image region: $\sum_{i = 1}^{n} \fkM_i = 0,$
\item $D_n~(n \geq 4)$ obeys the magnification relation in the $n$-image region: $\sum_{i = 1}^{n} \fkM_i = 0,$
\item $E_6$ obeys the magnification relation in the six-image region: $\sum_{i = 1}^{6} \fkM_i = 0,$
\item $E_7$ obeys the magnification relation in the seven-image region: $\sum_{i = 1}^{7} \fkM_i = 0,$
\item $E_8$ obeys the magnification relation in the eight-image region: $\sum_{i = 1}^{8} \fkM_i = 0.$
\end{enumerate}
In addition, for the two smooth generic three-parameter families of time delay functions $T_{c,\bby}$ {\rm(}or
induced lensing maps ${\boldsymbol \eta}_c${\rm)} in Table~\ref{table2}, and for any non-caustic target point  $\bs$ in the indicated region, the following results hold for the magnification $\mu_i \equiv \mu(\bx_i;\bs)$:
\begin{enumerate}
\item $D_{4}^{-}$ {\rm (Elliptic Umbilic)} Magnification relation in four-image region: 
$
\mu_{1} + \mu_{2} + \mu_{3} + \mu_{4}=0.
$
\item $D_{4}^{+}$ {\rm (Hyperbolic Umbilic)} Magnification relation in four-image region: 
$
\mu_{1} + \mu_{2} + \mu_{3} + \mu_{4}= 0.
$
\end{enumerate}
\end{theorem}
\vskip 10 pt
\noindent
{\it Remarks.} Note that for $n \geq 6$ there are Lagrangian maps that cannot be approximated by stable Lagrangian map-germs \cite{Arnold73}.

\section{Conclusion}
\label{Conclusion}
This paper provides a geometric explanation for the existence of magnification relations established in previous work.  In particular, it was established recently that for families of general mappings between planes exhibiting any of the $A_n~(n \geq 2), D_n~(n \geq 4), E_6, E_7, E_8$ family of caustic singularities, and for any non-caustic target point, the total signed magnification of the corresponding pre-images vanishes.  An identical result was also shown in the context of gravitational lensing, for lensing maps near elliptic umbilic $(D_4^{-})$ and hyperbolic umbilic $(D_4^{+})$ caustics.  A geometric reason for this fact comes to light when these mappings are extended to weighted projective space, which is a compact orbifold.  The orbifold version of the Global Residue Theorem then relates the total signed magnification to possible residues at infinity, which for the weighted projective spaces considered, do not in fact exist.  One then immediately concludes that the total signed magnification must be identically zero.  Thus magnification relations are transformed into a statement about the behavior of these mappings at infinity.  Our work generalizes multi-dimensional residue techniques in gravitational lensing developed in previous work.

\section{Acknowledgments}
\noindent ABA would particularly like to thank Paul Aspinwall and Dorette Pronk for helpful discussions on orbifolds.  AOP acknowledges the support of NSF Grant DMS-0707003. JMR thanks Mark Gross for the suggestion to consider weighted projective spaces.

\appendix
\section{Weighted Projective Space as a Compact Orbifold}
\label{WP}
In this section we provide a brief overview of orbifolds and of weighted projective space in particular, for the benefit of readers who may be unfamiliar with them.  Complex projective $n$-space $\CP^n$ is the set of 1-dimensional complex-linear subspaces of $\C^{n+1}$, with smooth quotient map $\pi\colon\,\C^{n+1}~\backslash~\{0\} \lra \CP^n$.  It is compact because the restriction of $\pi$ to the compact embedded submanifold $\SP^{2n+1} \subset \C^{n+1}$ is surjective.  We can also view $\CP^n$ as being obtained by the following smooth action of $\SP^1 \subset \C$ on $\SP^{2n+1}$:
\beq
\label{action1}
z \cdot (w_0,\dots,w_n) = (zw_0,\dots,zw_n).
\eeq
This action is {\it proper}.  This means by definition that the map $\rho\colon\,\SP^1 \times \SP^{2n+1} \lra \SP^{2n+1} \times \SP^{2n+1}$ defined by $z \cdot (w_0,\dots,w_n) = ((zw_0,\dots,zw_n),(w_0,\dots,w_n))$ is proper; i.e., for any compact set $K \subset \SP^{2n+1} \times \SP^{2n+1}$, its pre-image $\rho^{-1}(K) \subset \SP^1 \times \SP^{2n+1}$ is compact. Smooth actions are automatically proper if the Lie group is compact, as with $\SP^1$.  The action in \eqn{action1} is also {\it free}, because the stabilizer group
$$
\SP^1_{w} \equiv \{z \in \SP^1\,:\,z\cdot w = w\} = \{1\}
$$
for every $w \in \SP^{2n+1}$.  Being smooth, proper, and free guarantees that the resulting quotient space $\SP^{2n+1}/\SP^1$ is a smooth manifold, which is clearly diffeomorphic to $\CP^n$ (see, e.g., Lee 2006 \cite[Chap.~9]{Lee}).

Now consider generalizing the action defined by \eqn{action1}, as follows:
\beq
\label{action2}
z \cdot (w_0,\dots,w_n) = (z^{a_0}w_0,\dots,z^{a_n}w_n),
\eeq
where the $a_i$ are coprime positive integers.  This action is still smooth and proper, but it is no longer free: elements  in $\SP^{2n+1}$ of the form $(0,\dots,0,w_i,0,\dots,0)$ have stabilizer groups isomorphic to $\mathbb{Z}/a_i\mathbb{Z}$, because they are fixed by $a_i$th roots of unity.  Thus the action defined by \eqn{action2} is {\it almost free:} although the stabilizer group $\SP^1_{w}$ is not necessarily trivial for every $w \in \SP^{2n+1}$, it is always {\it finite}.  The resulting quotient space $\SP^{2n+1}/\SP^1 \equiv \WP(a_0,\dots,a_n)$ is known as {\it weighted projective space}, and it is not in general a manifold.  It is an example of an {\it orbifold,} which we now define.   Consult Satake 1956 \cite{Satake}, Moerdijk \& Pronk 1997 \cite{Moerdijk}, and \cite[Chap.~1]{Orbistrings} for a more detailed discussion of the material presented here.  For the algebro-geometric aspects of orbifolds, consult Beltrametti \& Robbiano 1986 \cite{Beltrametti} and Iano-Fletcher 2000 \cite{Iano-Fletcher}.

Let $X$ be a paracompact Hausdorff space, and define the following:
\begin{enumerate}
\item An {\it n-dimensional orbifold chart} is a connected open subset $\wt{U} \subset \RR^n$ and a continuous mapping $\phi\colon\,\wt{U} \lra \phi(\wt{U}) \equiv U \subset X$, together with a finite group $G$ of smooth automorphisms of $\wt{U}$ that satisfies the following condition: $\phi$ is $G$-invariant ($\phi \circ g = \phi$ for all $g \in G$) and induces a homeomorphism $\wt{U}/G \cong U$.  Let us clarify two points about this definition:

\begin{enumerate}
\item Here $\wt{U}/G$ is the quotient space defined by the usual quotient map $\pi\colon\,\wt{U} \lra \wt{U}/G$ (which is an open map because each $g\colon\,\wt{U}\lra \wt{U}$ is a diffeomorphism).  Because $\phi$ is $G$-invariant, it is constant on the fibers of $\pi$, so by the universal property of quotient maps it induces a unique continuous map $\varphi\colon\,\wt{U}/G \lra U$ satisfying $\varphi \circ \pi = \phi$.  In the definition of an orbifold chart we are therefore assuming that this map is a homeomorphism.

\item Any finite group is a compact zero-dimensional Lie group.  $G$ acts smoothly on $\wt{U}$ by hypothesis, and the action is proper because $G$ is compact.  If the action were also free, which we are {\it not} assuming, then the quotient space $\wt{U}/G$ would be a smooth manifold and the quotient map $\pi$ a smooth submersion.
\end{enumerate}
We write an orbifold chart as $(\wt{U},G,\phi)$.

\item Given two such charts $(\wt{U},G,\phi)$ and $(\wt{V},H,\psi)$, an {\it embedding} between them is a smooth embedding $\lambda\colon\,\wt{U} \longhookrightarrow \wt{V}$ satisfying $\psi \circ \lambda = \phi$.

\item Two orbifold charts $(\wt{U},G,\phi)$ and $(\wt{V},H,\psi)$ are {\it locally compatible} if every point $x \in U \cap V \subset X$ has an open neighborhood $W \subset U \cap V$ and an orbifold chart $(\wt{W},K,\mu)$ with embeddings $(\wt{W},K,\mu) \longhookrightarrow (\wt{U},G,\phi)$ and $(\wt{W},K,\mu) \longhookrightarrow (\wt{V},H,\psi)$.
\end{enumerate}
We say that $X$ is an {\it n-dimensional orbifold} if it has a maximal atlas of locally compatible $n$-dimensional orbifold charts.  Thus we see that an orbifold is locally modeled on quotients of open subsets of $\RR^n$ by finite group actions, and not simply open subsets of $\RR^n$ as with manifolds.  In general, therefore, orbifolds are not manifolds, though of course all manifolds are orbifolds.  However, if the finite group actions on the orbifold charts are all free, then $X$ is a smooth manifold by (1) and (3).

Now let $X$ be an orbifold.  For any $x \in X$, pick an orbifold chart $(\wt{U},G,\phi)$ containing it and pick a point $y$ in the fiber $\phi^{-1}(x) \subset \wt{U} \subset \RR^n$.  Define the {\it local group of x at y} to be
$$
G_y = \{g \in G : g(y) = y\}.
$$
If we instead choose another point $y' \in \phi^{-1}(x)$, then by (1a) above there is a (not necessarily unique) $g \in G$ such that $g(y) = y'$, and thus $G_{y'} = gG_{y}g^{-1}$.  If $(\wt{V},H,\psi)$ is another orbifold chart containing $x$, and if $\tilde{y} \in \psi^{-1}(x) \subset \wt{V}$ is any point in its fiber, then in fact $H_{\tilde{y}}$ and $G_y$ are also conjugate to each other (this fact is not trivial; it follows from the fact that an orbifold embedding $(\wt{U},G,\phi) \longhookrightarrow (\wt{V},H,\psi)$ induces an injective group homomorphism $G \longhookrightarrow H$; see \cite{Moerdijk}).  Thus the local group of $x$, which we now denote simply by $G_x$, is uniquely determined up to conjugacy.  If $G_x = 1$, then $x$ is said to be {\it regular;} if $G_x \neq 1$, then it is {\it singular}.  If $X$ has no singular points, then the local actions are all free, so $X$ is a smooth manifold.

The most common types of orbifolds are those that arise as quotient of manifolds by compact Lie groups.  In particular, if a compact Lie group $G$ acts smoothly, effectively (an action is {\it effective} if $g \cdot p = p$ for all $p \in M$ implies that $g = 1$), and almost freely on a smooth manifold $M$, then it can be shown that the resulting quotient space $M/G$ will be an orbifold as defined above; see \cite{Moerdijk,Orbistrings}, as well as Thurston 1980 \cite[Chap.~13]{Thurston}, for the details.  In particular, weighted projective space $\WP(a_0,\dots,a_n) \equiv \SP^{2n+1}/\SP^1$, with the action defined by \eqn{action2}, is a $2n$-dimensional orbifold. 

\section{Extending the Mappings in Tables~\ref{table1}~and~\ref{table2} to Weighted Projective Space}
\label{proof}

\subsection{Type $A_n\ (n \geq 2)$}
\label{typeA}

\noindent We begin with type $A_n$, $n \geq 2$.  The $(n-1)$-parameter family of general functions $F^{A_n}$ is given in \cite{Arnold73} by
\beq
\label{An1}
F^{A_n}(x,y) = \pm x^{n+1} \pm y^2 +c_{n-1}x^{n-1} + c_{n-2}x^{n-2} + \cdots + c_3x^3 + c_2x^2 + c_1x.
\eeq
To convert this into the form shown in Table~\ref{table1}, we use the following coordinate transformation on the domain $\{(x,y)\} = \RR^2$:
\beq
\label{An_coord}
(x,y) \longmapsto \left(x,y+\frac{s_2}{2}\right).\nonumber
\eeq
This transforms \eqn{An1} to
\beq
\label{An}
F_{\bc,\bs}^{A_n}(x,y) = \pm x^{n+1} \pm y^2 + c_{n-1}x^{n-1} + c_{n-2}x^{n-2} + \cdots + c_3x^3 + s_2x^2 - s_1x \pm s_2y,
\eeq
where $c_1 \equiv -s_1$ and $c_2 \equiv s_2$.  The parameters $s_1, s_2$ are to be interpreted in the context of gravitational lensing as the rectangular coordinates on the source plane $S = \RR^2$.  Note that we omitted the constant term from \eqn{An} since it will not affect any of our results below.  Note also that
$$
\text{det\big(Hess}\,F^{A_n}\big) = \text{det\big(Hess}\,F_{\bc,\bs}^{A_n}\big),
$$
so that the magnification (as defined in eqn.~\eqref{Gauss}) is unaltered.  Henceforth we will work with the form of $F_{\bc,\bs}^{A_n}$ in \eqn{An}.  To derive the induced mapping $\fA$, we proceed as follows.  For the fold singularity $(A_2)$, for example, we set $n=2$ in \eqn{An} and equate the partial derivatives $\partial F_{\bc,\bs}^{A_2}/\partial x$ and $\partial F_{\bc,\bs}^{A_2}/\partial y$ to zero:
\beq
\frac{\partial F_{\bc,\bs}^{A_2}}{\partial x} = \pm3x^2 - s_1 = 0\hspace{.25 in},\hspace{.25 in}
\frac{\partial F_{\bc,\bs}^{A_2}}{\partial y} = \pm2y \pm s_2 = 0.\nonumber
\eeq
Solving for the coordinates $s_1, s_2$ gives us ${\bbf}^{A_2}$:
\beq
\label{fold}
{\bbf}^{A_2}(x,y) = (\pm3x^2,-2y) = (s_1,s_2).
\eeq
The same procedure for any $n$ leads to the following family of general mappings $\fA$:
\beq
\label{Anmap2}
\fA(x,y) = \left(\pm(n+1)x^{n} + (n-1)c_{n-1}x^{n-2} + \cdots + 3c_3x^2 - 4xy~,\,-2y\right) = (s_1,s_2),
\eeq
whose pre-images $(x_i,y_i)$ have magnification $\left(\detfA\right)^{-1}(x_i,y_i) = \text{det\big(Hess}\,F_{\bc,\bs}^{A_n}\big)^{-1}(x_i,y_i) \equiv \fkM_i$.  The simple form of the leading order terms in \eqn{Anmap2} suggests that we extend $\fA$ to $\mathbb{WP}(1,1,1) = \mathbb{CP}^2$.  Indeed, in homogeneous coordinates $[X:Y:U]$, the solutions of \eqn{Anmap2} are the common roots in affine space $(U = 1)$ of the following two homogeneous polynomials in $\mathbb{WP}(1,1,1)$: 
\beq
\text{${\bbf}^{A_n}$ homogenized in $\mathbb{WP}(1,1,1)~\Longrightarrow$}~~
\left\{
\begin{array}{ll}
\pm(n+1)X^{n} + (n-1)c_{n-1}X^{n-2}U^2 + \cdots + 3c_3X^2U^{n-2} - 4XYU^{n-2} - s_1U^n\nonumber \\
-2Y - s_2U.\nonumber
\end{array}
\right.
\eeq
The common roots at infinity are obtained by setting $U = 0$, which yields only the root $[0:0:0] \notin \WP(1,1,1)$.  Moreover, since $\mathbb{WP}(1,1,1) = \mathbb{CP}^2$ is a (compact) smooth manifold, it has no singular points.  The Global Residue Theorem then tells us that the sum of the residues in affine space is minus the sum of the residues at infinity.  Since there are no residues at infinity, the magnification theorem immediately follows.

\subsection{Type $D_n\ (n \geq 4)$}
\label{typeD}

\noindent For type $D_n$, $n \geq 4$, the corresponding $(n-3)$-parameter family of induced general maps $\fD$ is shown in Table~\ref{table1}:
\beq
\label{Dnmap2}
\fD(x,y) = \left(2xy~,~x^2\pm(n-1)y^{n-2} + (n-2)c_{n-2}y^{n-3} + \cdots + 2c_2y\right) = (s_1,s_2).
\eeq
We now extend \eqn{Dnmap2} to the weighted projective space $\mathbb{WP}(n-2,2,1)$, so that the affine coordinates $x,y$ are related to the homogeneous coordinates $[X:Y:U]$ by
\beq
x = \frac{X}{U^{n-2}}\hspace{.25 in},\hspace{.25 in}y = \frac{Y}{U^2}\cdot
\eeq
The solutions of \eqn{Dnmap2} are the common roots in affine space $(U = 1)$ of the following two homogeneous polynomials in $\mathbb{WP}(n-2,2,1)$:
\beq
\text{$\fD$ homogenized in $\mathbb{WP}(n-2,2,1)~\Longrightarrow$}~~
\left\{
\begin{array}{ll}
2XY - s_1U^{n}\nonumber\\
X^2 \pm (n-1)Y^{n-2} + (n-2)c_{n-2}Y^{n-3}U^2 + \cdots + 2c_2YU^{2n-6} - s_2U^{2n-4}.\nonumber
\end{array}
\right.
\eeq
(Note that these polynomials {\it are} homogeneous, since $X$ and $Y$ now have weights $n-2$ and $2$, respectively; the degree of the term $2XY$, for example, is $(n-2) + 2 = n$.)  At infinity $(U=0)$, the only common root is $[0:0:0] \notin \mathbb{WP}(n-2,2,1)$.  Note that the singular points of $\WP(n-2,2,1)$ occur at infinity.

\subsection{Type $E_6$}
\label{typeE6}

\noindent The 5-parameter family of induced general maps $\fEsix$ corresponding to type $E_6$ is shown in Table~\ref{table1}:
\beq
\label{E6map}
\fEsix(x,y) = \left(3x^2 + c_3y^2 +c_1y~,~\pm4y^3 + 2c_3xy + 2c_2y + c_1x\right) = (s_1,s_2).
\eeq
As with $A_n$, we can extend \eqn{E6map} to $\mathbb{WP}(1,1,1) = \mathbb{CP}^2$, with corresponding homogeneous polynomials
\beq
\text{$\fEsix$ homogenized in $\mathbb{WP}(1,1,1)~\Longrightarrow$}~~
\left\{
\begin{array}{ll}
3X^2 + c_3Y^2 +c_1YU - s_1U^2\nonumber\\
\pm4Y^3 +2c_3XYU +2c_2YU^2 + c_1XU^2 - s_2U^3.\nonumber
\end{array}
\right.
\eeq
The only common root at infinity $(U = 0)$ is $[0:0:0] \notin \mathbb{WP}(1,1,1)$.

\subsection{Type $E_7$}
\label{typeE7}

\noindent For type $E_7$, Table~\ref{table1} gives the corresponding 4-parameter family of induced general maps $\fEseven$:
\beq
\label{E7map2}
\fEseven(x,y) = \left(3x^2 + y^3 + c_1y~,~3xy^2 + 4c_4y^3 + 3c_3y^2 + 2c_2y + c_1x\right) = (s_1,s_2).
\eeq
We extend \eqn{E7map2} to $\mathbb{WP}(3,2,1)$, with homogeneous coordinates
\beq
x = \frac{X}{U^{3}}\hspace{.25 in},\hspace{.25 in}y = \frac{Y}{U^2}\nonumber
\eeq
and corresponding homogeneous polynomials
\beq
\text{$\fEseven$ homogenized in $\mathbb{WP}(3,2,1)~\Longrightarrow$}~~
\left\{
\begin{array}{ll}
3X^2 + Y^3 +c_1YU^4 - s_1U^6\nonumber\\
3XY^2 + 4c_4Y^3U + 3c_3Y^2U^3 + 2c_2YU^5 + c_1XU^4 - s_2U^7,\nonumber
\end{array}
\right.
\eeq
whose common roots in affine space $(U=1)$ are precisely the solutions to \eqn{E7map2}.  The only common root at infinity $(U = 0)$ is $[0:0:0] \notin \mathbb{WP}(3,2,1)$.

The polynomials in this case have weighted degrees 6 and 7.
As an example of the vanishing criterion quoted in \eqn{neg-deg}, a form $\omega$ as in
\eqn{omega-h} would have vanishing sum of residues in $\C^2$ when $\deg h < 6+7-3-2=8$.

\subsection{Type $E_8$}
\label{typeE8}

\noindent Finally, Table~\ref{table1} gives the 5-parameter family of induced general mappings $\fEeight$ corresponding to type $E_8$:
\beq
\label{E8map}
\fEeight(x,y) = \left(3x^2 + c_5y^3 + c_4y^2 + c_1y~,~5y^4 + 3c_5xy^2 + 2c_4xy + 3c_3y^2 + 2c_2y + c_1x\right) = (s_1,s_2).\nonumber
\eeq
Once again we will use $\mathbb{WP}(3,2,1)$.  This time the corresponding homogeneous polynomials are
\beq
\text{$\fEeight$ homogenized in $\mathbb{WP}(3,2,1)~\Longrightarrow$}~~
\left\{
\begin{array}{ll}
3X^2 + c_5Y^3 + c_4Y^2U^2 + c_1YU^4 - s_1U^6\nonumber\\
5Y^4 + 3c_5XY^2U + 2c_4XYU^3 + 3c_3Y^2U^4 + 2c_2YU^6 + c_1XU^5 - s_2U^8.\nonumber
\end{array}
\right.
\eeq
And, of course, the only common root at infinity $(U = 0)$ is $[0:0:0]$.

\subsection{Quantitative Forms for the Elliptic Umbilic and Hyperbolic Umbilic}
\label{Quantitative}

\noindent {\bf Elliptic and Hyperbolic Umbilics:} Table~\ref{table2} gives the universal, quantitative form of a lensing map in the neighborhood of either an elliptic umbilic $(D_4^{-})$ or hyperbolic umbilic $(D_4^{+})$ caustic.  These are both one-parameter maps $\blm_c$ induced by a three-parameter time delay family $T_{c,\bs}$.  (In the context of gravitational lensing, $c$ will represent some physical input, such as the source redshift.)  For the elliptic umbilic, this induced mapping is
$$
\blm_{c,\text{ell}}(x,y) = \left(x^2-y^2~,\,-2xy+4cy\right) = (s_1,s_2),
$$ 
while for the hyperbolic umbilic, it is
$$
\blm_{c,\text{hyp}}(x,y) = \left(x^2+2cy~,\,y^2+2cx\right) = (s_1,s_2).
$$ 
The desired extension in both cases is to $\WP(1,1,1) = \CP^2$, with corresponding homogeneous polynomials
\beq
\text{$\blm_{c,\text{ell}}$ homogenized in $\mathbb{WP}(1,1,1)~\Longrightarrow$}~~
\left\{
\begin{array}{ll}
X^2 - Y^2 -s_1U^2\nonumber\\
-2XY + 4cYU - s_2U^2\nonumber
\end{array}
\right.
\eeq
and

\beq
\hspace{-.23in}\text{$\blm_{c,\text{hyp}}$ homogenized in $\mathbb{WP}(1,1,1)~\Longrightarrow$}~~
\left\{
\begin{array}{ll}
X^2 + 2cYU - s_1U^2\nonumber\\
Y^2 + 2cXU - s_2U^2.\nonumber
\end{array}
\right.
\eeq
There are no common roots at infinity $(U=0)$ in either case.~~~$\square$



\begin{thebibliography}{}

\bibitem{Aazami-Petters}
Aazami, A. B., Petters, A. O.,
{\it J. Math. Phys.} (2009), {\bf 50}, 032501.

\bibitem{Aazami-Petters2}
Aazami, A. B., Petters, A. O.,
{\it J. Math. Phys.} (2009), {\bf 50}, 082501.

\bibitem{Aazami-Petters3}
Aazami, A. B., Petters, A. O.,
{\it J. Math. Phys.} (2010), {\bf 51}, 023503.

\bibitem{Orbistrings}
Adem, A., Leida, J., and Ruan, Y.,
{\it Orbifolds and Stringy Topology} (CUP, 2007).

\bibitem{Arnold73}
Arnold, V. I.,
{\it Func. Anal. Appl.} (1973), {\bf 6}, 254.

\bibitem{AGV1}
Arnold, V. I., Gusein-Zade, S. M., and Varchenko, A. N.,
{\it Singularities of Differentiable Maps, vol. I} (Birkh{\"a}user, Boston, 1985).

\bibitem{Beltrametti}
Beltrametti, M., Robbiano, L.,
{\it Expositiones Mathematicae} (1986), {\bf 4}, 111.

\bibitem{Blandford90}
Blandford, R. D.,
{\it Q. Jl. Roy. Astron. Soc.} (1990), {\bf 31}, 305.

\bibitem{Blan-Nar}
Blandford, R. D., Narayan, R.,
{\it Astrophys. J.} (1986), {\bf 310}, 568.


\bibitem{C-Hayes}
Castrigiano, D., Hayes, S.,
{\it Catastrophe Theory} (Westview, Boulder, 2004).

\bibitem{Cattani-Cox}
Cattani, E., Cox, D., and Dickenstein, A., {\it Compositio Mathematica} (1997) {\bf 108}, 35.

\bibitem{Cattani-Dickenstein}
Cattani, E., Dickenstein, A., and Sturmfels, B., in {\it Algorithms in Algebraic Geometry and Applications}, 
p. 135, ed. L. Gonzalez-Vega \& T. Recio (Birkh{\"a}user, Basel, 1996).

\bibitem{Dalal-Rabin}
Dalal, N., Rabin, J. M.,
{\it J. Math. Phys.} (2001), {\bf 42}, 1818.


\bibitem{Gol-G}
Golubitsky, M., Guillemin, V.,
{\it Stable Mappings and Their Singularities} (Springer, Berlin, 1973).

\bibitem{Griffiths-Harris}
Griffiths, P., Harris, J.,
{\it Principles of Algebraic Geometry} (Wiley, New York, 1978).

\bibitem{Iano-Fletcher}
Iano-Fletcher, A. R.,
{\it London Math. Soc. Lecture Note Ser.} (2000), {\bf 281}, 101.

\bibitem{KGP-cusps}
Keeton, C., Gaudi, S., and Petters, A.O.,
{\it Astrophys. J.} (2003), {\bf 598}, 138.

\bibitem{KGP-folds}
Keeton, C., Gaudi, S., and Petters, A.O.,
{\it Astrophys. J.} (2005), {\bf 635}, 35.

\bibitem{Lee}
Lee, J. M.,
{\it Introduction to Smooth Manifolds} (Springer, New York, 2006).

\bibitem{Majthay}
Majthay, A., 
{\it Foundations of Catastrophe Theory} (Pitman, Boston, 1985).

\bibitem{Mao-Sch}
Mao, S., Schneider, P.,
{\it Mon. Not. Roy. Astron. Soc.} (1998), {\bf 295}, 587.

\bibitem{Moerdijk}
Moerdijk, I., Pronk, D. A.,
{\it K-Theory} (1997), {\bf 12}, 3.

\bibitem{Petters93}
Petters, A. O.,
{\it J. Math. Phys.} (1993), {\bf 34}, 3555.

\bibitem{Petters}
Petters, A. O., Levine, H., and Wambsganss, J., 
{\it Singularity Theory and Gravitational Lensing} (Birkh{\"a}user, Boston, 2001).

\bibitem{Satake}
Satake, I.,
{\it Proc. Nat. Acad. Sci. USA} (1956), {\bf 42}, 359.

\bibitem{Sch-EF}
Schneider, P., Ehlers, J., and Falco, E., 
{\it Gravitational Lenses} (Springer, Berlin, 1992).

\bibitem{Sch-Weiss92}
Schneider, P., Weiss, A.,
{\it Astron. Astrophys.} (1992), {\bf 260}, 1.

\bibitem{Thurston}
Thurston, W.,
{\it The Geometry and Topology of Three-Manifolds}, Princeton Lecture Notes, 1980, unpublished.

\bibitem{Werner}
Werner, M. C.,
{\it J. Math. Phys.} (2007), {\bf 48}, 052501.

\bibitem{Werner2}
Werner, M. C.,
{\it J. Math. Phys.} (2009), {\bf 50}, 082504.

\bibitem{Zakharov}
Zakharov, A.,
{\it Astron. Astrophys.} (1995), {\bf 293}, 1.

\end{thebibliography}
\end{document}